\documentclass[
    aps,
    prl,
    letterpaper,
    10pt,
    reprint,
    twocolumn,
    nofootinbib,
    longbibliography,
    amsmath,amssymb,amssymb,
    floatfix,
    superscriptaddress,
    noeprint
]{revtex4-2}

\usepackage[english]{babel}
\usepackage[utf8]{inputenc}
\usepackage[T1]{fontenc}
\usepackage{graphicx}
\usepackage{dcolumn}
\usepackage{bbm}
\usepackage{bm}
\usepackage{braket}
\usepackage{comment}
\usepackage[dvipsnames]{xcolor}
\usepackage[colorlinks,%
            linkcolor=BrickRed,%
            citecolor=MidnightBlue,%
            urlcolor=MidnightBlue]{hyperref}
\usepackage[normalem]{ulem}
\usepackage{amsmath,amssymb}
            
\newcounter{ga}

\newcounter{ls}

\newcounter{pr}

\newcounter{mh}

\begin{document}
\title{Dissipation- versus Chaos-Induced Relaxation\\in Non-Markovian Quantum Many-Body Systems}%

\author{Gabriel Almeida}
\email{gabriel.m.almeida@tecnico.ulisboa.pt}
\affiliation{CeFEMA-LaPMET, Departamento de Física, Instituto Superior Técnico, Universidade de Lisboa, Av. Rovisco Pais, 1049-001 Lisboa, Portugal}

\author{Pedro Ribeiro}
\email{ribeiro.pedro@tecnico.ulisboa.pt}
\affiliation{CeFEMA-LaPMET, Departamento de Física, Instituto Superior Técnico, Universidade de Lisboa, Av. Rovisco Pais, 1049-001 Lisboa, Portugal}
\affiliation{Beijing Computational Science Research Center, Beijing 100193, China}

\author{Masudul Haque}
\affiliation{Institut f\"ur Theoretische Physik, Technische Universit\"at Dresden, 01062 Dresden, Germany}

\author{Lucas  S\'a}
\email{ld710@cam.ac.uk}
\affiliation{TCM Group, Cavendish Laboratory, University of Cambridge, JJ Thomson Avenue, Cambridge, CB3 0US, UK\looseness=-1}

\begin{abstract}
In interacting quantum many-body systems, relaxation toward equilibrium reflects a competition between internal chaotic dynamics and environmental dissipation. While conventional Markovian baths typically produce exponential decay, non-Markovian dissipation can give rise to more intricate behavior, including algebraic relaxation. We study an open Sachdev-Ye-Kitaev (SYK) model coupled to a pseudogapped fermionic bath, using the Keldysh formalism to compute steady-state correlations in the large-$N$ limit. Our results uncover a rich dynamical phase diagram, with regimes of bath-driven power-law relaxation, chaos-driven exponential decay, and an intermediate pre-relaxation phase where exponential decay crosses over to algebraic decay. These findings demonstrate that non-Markovian environments can qualitatively reshape relaxation mechanisms in strongly correlated quantum many-body systems.
\end{abstract}

\maketitle

Understanding the relaxation of interacting quantum many-body systems towards a steady state is a central problem in modern physics. For isolated systems, relaxation and thermalization are now understood in the framework of the eigenstate thermalization hypothesis (ETH) \cite{deutsch1991quantum,srednicki1994chaos,d2016quantum}, which is a manifestation of quantum chaos. 
Beyond establishing the mechanism of thermalization, recent works \cite{prosen2002,prosen2004,mori2024,yoshimura2024,znidaric2024,znidaric2024b,yoshimura2025,duh2025} have clarified how fast relaxation occurs.
In realistic settings, however, quantum systems are inevitably coupled to external degrees of freedom. Thus, the resulting dynamics features a competition between internal equilibration driven by quantum chaos and relaxation imposed by the environment. Most theoretical descriptions of open quantum systems rely on the Markovian approximation, assuming weak system-bath coupling and short environmental memory times \cite{davies1974markovian,breuer2002theory}. In this regime, the dynamics is governed by a Lindblad master equation \cite{gorini1976completely,lindblad1976generators}, and relaxation toward the steady state is typically exponential and controlled by the spectral gap of the corresponding Lindbladian \cite{znidaric2015,mori2020,mori2023,shirai2024,bao2026}. Deviations from exponential relaxation may nevertheless arise when the spectral gap closes \cite{kessler2012dissipative,poletti2012interacting,cai2013algebraic,znidaric2015,minganti2018spectral} or when the Liouvillian eigenmodes become localized \cite{znidaric2015,haga2021}.

This picture relies implicitly on an environment with a flat density of states. When the bath instead exhibits nontrivial structure, memory effects emerge and the Markovian description breaks down \cite{weiss2012quantum,rivas2014RPP,breuer2016colloquium,de2017dynamics}. The resulting non-Markovian dynamics has been extensively studied in quantum Brownian motion~\cite{jung1985longtime,grabert1988quantum}, two-level systems~\cite{leggett1987dynamics,garraway1997decay,dalton2003Nonmarkovian,mazzola2009pseudomodes,vacchini2010exact,Monteiro2025Nonmarkovian}, free fermions and bosons~\cite{chakraborty2018power} and impurities~\cite{arrigoni2020masterequation}, using methods ranging from quantum master equations to functional-integral approaches and the Keldysh formalism. In these systems, structured reservoirs are known to induce slow, nonexponential relaxation.

By contrast, the impact of structured environments on the relaxation of strongly correlated quantum systems remains largely unexplored. A particularly important class of non-Markovian reservoirs are pseudogapped baths, whose density of states vanishes as a power law at low frequencies \cite{withoff1990phase,vojta2002fractional}. Such environments naturally arise in platforms like graphene \cite{mathe2020nonequilibrium,minarelli2022two,PhysRevLett.133.126503} and engineered reservoirs \cite{li2020construction,PhysRevLett.102.166806,PhysRevB.81.113402}, making them a relevant setting to investigate how memory effects modify relaxation beyond the weakly interacting paradigm.

\begin{figure}
	\centering
	\includegraphics[width=0.99\linewidth,trim=1cm 2.2cm 14cm 2cm,clip]{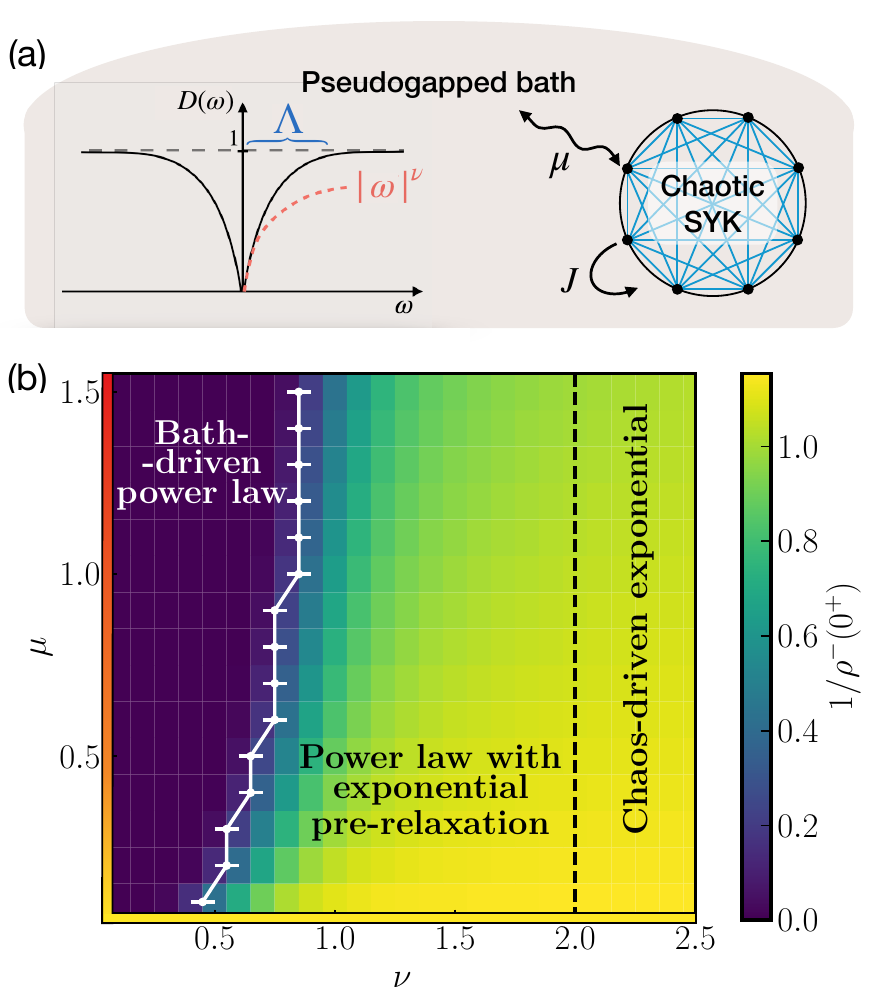}
	\caption{(a) Sketch of the model: an SYK system coupled to an environment with a pseudogapped density of states $D(\omega)$, characterized by the exponent $\nu$, pseudogap width $\Lambda$, and inverse temperature $\beta$. The dissipation, with strength $\mu$, competes with internal chaotic dynamics, with strength $J$, to equilibrate the system. 
	(b) Dynamical phase diagram in the $(\mu, \nu)$ plane for $J=1$, $\beta=0$, and $\Lambda=10$, obtained by combining the results of this Letter. It features a bath-driven power-law relaxation regime, a chaos-driven exponential relaxation regime, and an intermediate pre-relaxation regime where an initial exponential decay is followed by asymptotic algebraic relaxation. 
    Different phases are distinguished by the behavior of steady-state correlation functions [such as $\rho^-(\omega)$ defined in the text] at small frequencies, namely, by the existence of a divergence or nonanaliticity of $\rho^-(0^+)$. The white boundary separates the region where $1/\rho^-(0^+) < \delta = 0.02$ (signaling a divergence) and the error bars correspond to the spacing of the $\nu$-grid. The vertical dashed line separates the region of analytic and nonanalytic behavior of $\rho^-(0^+)$.
    The $\mu = 0$ stripe is highlighted in yellow, corresponding to the isolated system where relaxation is exponential and driven by chaos. The $\nu=0$ stripe is colored from yellow to red, since it corresponds to the Markovian limit, where relaxation is exponential with a crossover from bath-induced relaxation (red) at large $\mu$ to chaos-induced relaxation at small $\mu$ (yellow).
	}
	\label{fig:cartoon-phase-diagram}
\end{figure}

In this Letter, we investigate the relaxation dynamics of a strongly correlated model coupled to a pseudogapped fermionic bath; see Fig.~\ref{fig:cartoon-phase-diagram}(a). We find that the pseudogapped environment can induce power-law relaxation, whose exponent can be computed analytically. Furthermore, our results show that the competition between internal dynamics and structured dissipation leads to distinct dynamical phases, including exponential relaxation driven by quantum chaos, algebraic relaxation driven by the environment, and a pre-relaxation phase; see Fig.~\ref{fig:cartoon-phase-diagram}(b). We compute this dynamical phase diagram at infinite bath temperature and demonstrate its robustness at finite temperature in the Supplemental Material (SM)~\cite{SM}. 

To address this problem in a controlled way, it is useful to consider analytically tractable models that capture the essential features of quantum chaos and strong correlations. A paradigmatic example is the Sachdev-Ye-Kitaev (SYK) model \cite{sachdev1993gapless,kitaev2015,sachdev2015bekenstein}: a zero-dimensional model of Majorana fermions with random all-to-all interactions, whose large-$N$ limit is exactly solvable and displays maximal quantum chaos \cite{maldacena2016remarks}. 
In equilibrium, SYK realizes a non-Fermi liquid state with no quasiparticle poles, an extensive zero-temperature entropy, and an emergent approximate conformal symmetry governing its low-energy dynamics. Because of its analytical control and rich dynamical structure, SYK has become a powerful platform for studying relaxation in strongly correlated quantum systems, both in isolation \cite{eberlein2017quantum,bhattacharya2019quantum,PhysRevLett.124.106401,samui2021thermalization,larzul2022quenches,PhysRevB.105.075117,berenguer2024} and with an external bath, where dissipation emerges from the unitary dynamics of a combined system–bath Hamiltonian \cite{banerjee2017solvable,chen2017tunable,zhang2019evaporation,PhysRevB.102.224305,PhysRevResearch.2.013307,cheipesh2021quantum,Maldacena2021,PhysRevResearch.4.023001,Almheiri2024,berenguer2025,wang2026} (see also Refs.~\cite{Chen2021,Su2021,Zhang2023,zhang2025scrambling} for entropy dynamics, Refs.~\cite{delCampo2021thermofield,delCampo2022spectral} for energy dephased SYK models and Refs.~\cite{PhysRevLett.127.140601,jian2021quantum,PhysRevResearch.4.L022066,Jian2023,milekhin2024,gopalakrishnan2026tailoring} for monitored SYK models). 
Motivated by this, Lindbladian versions of the SYK model have been recently introduced \cite{sa2022lindbladian} to study the interplay between strong correlations and Markovian dissipation, including explicit computations of the spectral gap \cite{sa2022lindbladian,kulkarni2022lindbladian,garcia2023keldysh,zheng2025exceptional}, Lyapunov exponents \cite{garcia2024lyapunov,liu2026SYK}, dynamical phase transitions \cite{kawabata2023dynamical,wang2024,chen2025SWSSB}, operator size and Krylov complexity \cite{bhattacharjee2023operator,bhattacharjee2024operator,srivatsa2024,liu2024}, spectral statistics \cite{sa2022lindbladian,li2024}, symmetry classifications \cite{kawabata2023symmetry,garcia2024toward}, transient topological modes \cite{garcia2025topology}, quantum many-body scars \cite{garcia2026scars}, and Lindbladian ETH \cite{almeida2026}.

We consider an open SYK model coupled to a structured fermionic bath. The system consists of $N$ Majorana fermions, $\chi_i = \chi_i^\dagger$, that satisfy the Clifford algebra $\{\chi_i, \chi_j\} = 2\delta_{ij}$ and with Hamiltonian 
\begin{equation}
    H = \sum_{i<j<k<l}^N J_{ijkl} \chi_i \chi_j \chi_k \chi_l,
    \label{eq:syk-hamiltonian}
\end{equation}
where $J_{ijkl}$ are random couplings sampled from a Gaussian distribution with zero mean and variance $\langle J_{ijkl}^2 \rangle = 3! J^2/N^3$. 
Dissipation is introduced by coupling the system linearly to a fermionic environment held at inverse temperature $\beta$,
\begin{equation}
    H_\mathrm{SE} = \sqrt{\mu} \sum_{i}  \chi_i Y_i,
    \label{eq:coupling}
\end{equation}
where $Y_i$ are Majorana bath operators and $\mu$ controls the overall strength of the dissipation. The bath is assumed to remain in thermal equilibrium, so its correlation functions are time-translation invariant and fully characterized by the density of states $D(\omega)$.

We focus on environments with a pseudogapped density of states, $D(\omega)\sim|\omega|^\nu$ as $\omega \to 0$, with $\nu\geq 0$, which suppresses low-energy bath modes and naturally leads to non-Markovian dissipation. For concreteness, we use the smooth form
\begin{equation}
D(\omega)=\left(1-e^{-\omega^2/\Lambda^2}\right)^{\nu/2},
\label{eq:pseudogap-dos}
\end{equation}
where $\Lambda$ is the pseudogap width. $D(\omega)$ reproduces the desired low-frequency scaling, while saturating at high frequencies $\omega \gg \Lambda$. For $\nu = 0$, the density of states becomes constant, corresponding to a metallic environment. When combined with an infinite-temperature bath ($\beta = 0$), this limit reproduces Markovian dissipation since the memory-kernel becomes a $\delta$-function in time; this limit was studied in Ref. \cite{garcia2023keldysh}. In contrast, taking $\nu \to \infty$ yields a hard gap, $D(\omega) = \Theta(|\omega| - \Lambda)$, which suppresses all bath modes below the pseudogap width. Throughout this Letter, we set $J=1$ and take $\Lambda$ much larger than all other energy scales.

To characterize relaxation toward the steady state, we focus on the retarded Green’s function
\begin{equation}
    i G^R(t) = \Theta(t)\,
    \Bigl\langle \operatorname{Tr}\left[ \rho_\infty \{\chi_i(t), \chi_i(0)\} \right] \Bigr\rangle,
    \label{eq:Gr-def}
\end{equation}
where $\rho_\infty$ denotes the steady-state density matrix, and the bracket denotes an average over the disordered coefficients $J_{ijkl}$ and fermion indices. The long-time decay of $G^R(t)$ encodes the nature of the relaxation: for instance, whether correlations decay exponentially or algebraically, and which decay rates or exponents control this behavior. As such, determining $G^R(t)$ provides direct insight into the interplay between strong interactions and non-Markovian dissipation in open quantum systems.

We compute $G^R(t)$ using a Keldysh path-integral formulation appropriate for open quantum systems \cite{keldysh,Sieberer_2016,kamenev2023field,reyes2026schwinger}. After integrating out the bath degrees of freedom and performing the disorder average, we find an effective action for collective fields of the system. In the large-$N$ limit, only the saddle point of the action contributes, which yields a closed set of Schwinger-Dyson equations for the steady-state correlation functions; see the SM \cite{SM}.
These include the spectral function $\rho^-(\omega)$, obeying the sum rule
$\int d\omega\, \rho^-(\omega) = 1$, which together with its Hilbert transform $\rho^H(\omega)$ fully determines the retarded Green’s function in frequency space,
\begin{equation}
    G^R(\omega) = -\pi \left[\rho^H(\omega) + i \rho^-(\omega)\right].
    \label{eq:GR(omega)}
\end{equation}
Fourier transforming and noting that $\rho^-(\omega)$ is even gives
\begin{equation}
    i G^R(t)
    = \Theta(t) \int_{-\infty}^{+\infty} d\omega\,
    \rho^-(\omega)\, \cos(\omega t).
    \label{eq:Gr-from-rho}
\end{equation}
Hence, the low-frequency behavior of $\rho^-(\omega)$ directly controls the late-time relaxation of the system towards its steady state.
Another relevant quantity is the fluctuation function $\rho^+(\omega)$, proportional to the Keldysh component of the Green’s function. 

\begin{figure}
    \centering
    \includegraphics[width=0.99\linewidth]{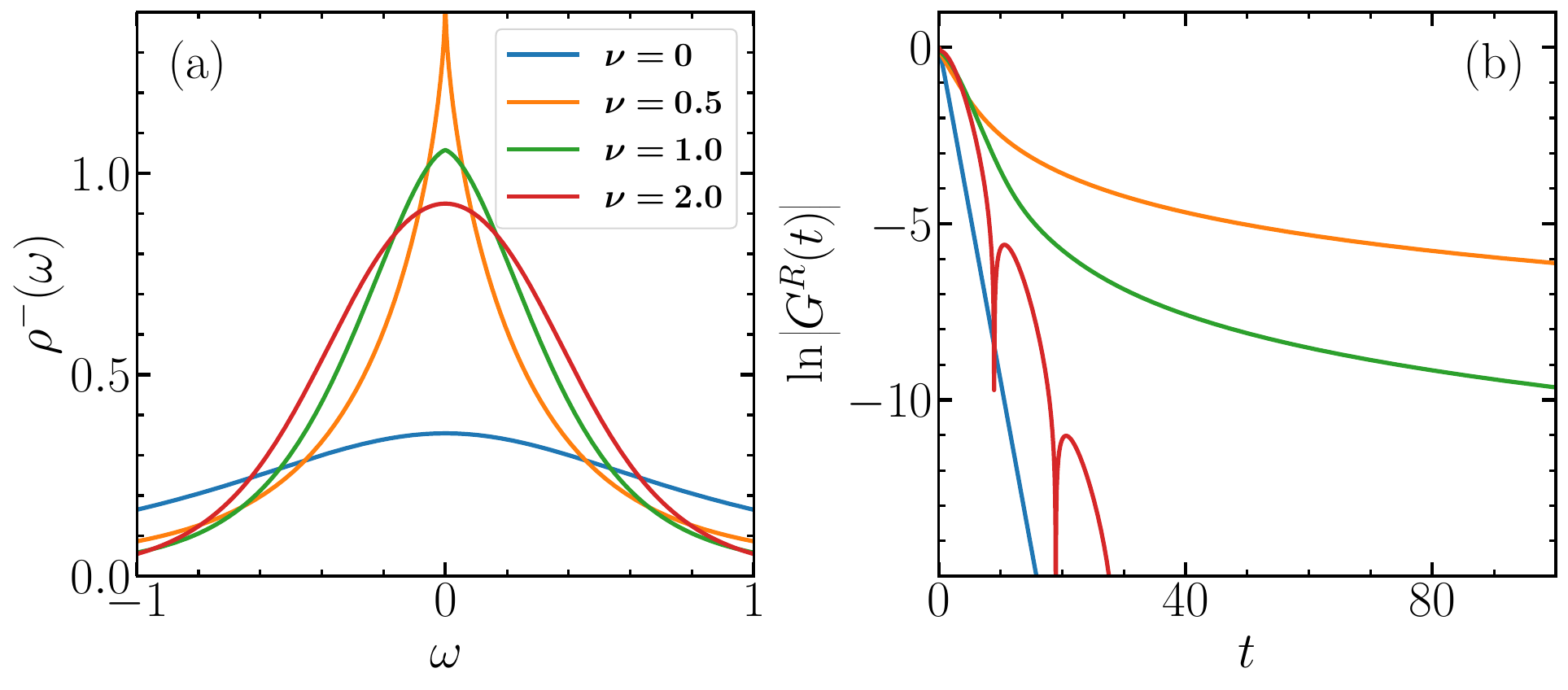}
    \caption{Solution of the Schwinger-Dyson equations for $\beta=0$, $\Lambda=10$, $\mu=0.8$, and several values of the pseudogap exponent $\nu$. (a) Depending on $\nu$, the spectral function $\rho^-(\omega)$ is either Lorentzian-shaped or develops a nonanalytic cusp at $\omega=0$. (b) Correspondingly, the retarded Green’s function $G^R(t)$ exhibits either exponential or power-law decay in time.
    }
    \label{fig:some-solutions}
\end{figure}

In equilibrium at inverse temperature $\beta$, the correlation functions satisfy the fluctuation-dissipation relation
\begin{equation}
\rho^+(\omega) = \rho^-(\omega) \tanh\left(\frac{\beta \omega}{2}\right),
\label{eq:fdr-rhos}
\end{equation}
which simplifies the Schwinger-Dyson equations. For an infinite-temperature bath ($\beta = 0$), they reduce to (see the SM \cite{SM} for the finite-$\beta$ equations)
\begin{align}
    &\sigma^-(\omega) = \frac{J^2}{4} \left(\rho^- \ast \rho^- \ast \rho^- \right)(\omega) + \frac{\mu}{\pi} D(\omega), \label{eq:sde-sigma-minus}
    \\
    &\rho^- (\omega) = \frac{\sigma^- (\omega)}{(\omega  + \pi \sigma^H(\omega))^2 + (\pi \sigma^-(\omega))^2},
    \label{eq:sde-rho}
\end{align}
where $\sigma^-$ is the self-energy, $\sigma^H$ its Hilbert transform and $\ast$ denotes a convolution. In the Markovian limit ($\nu=0$), Eqs.~\eqref{eq:sde-sigma-minus} and \eqref{eq:sde-rho} reduce to those studied in Ref.~\cite{garcia2023keldysh}, which found exponential relaxation, providing a benchmark for our results. 
We solved Eqs.~(\ref{eq:sde-sigma-minus}) and (\ref{eq:sde-rho}) self-consistently \cite{ribeiro2015steadystate,sa2022lindbladian} for representative parameters (see the SM \cite{SM} for details).
The resulting solutions, shown in Fig.~\ref{fig:some-solutions}, exhibit either exponential or power-law relaxation depending on the pseudogap exponent $\nu$. We explain this behaviour in the following by analyzing the low-frequency structure of $\rho^-(\omega)$.

\begin{figure}
    \centering
    \includegraphics[width=0.99\linewidth]{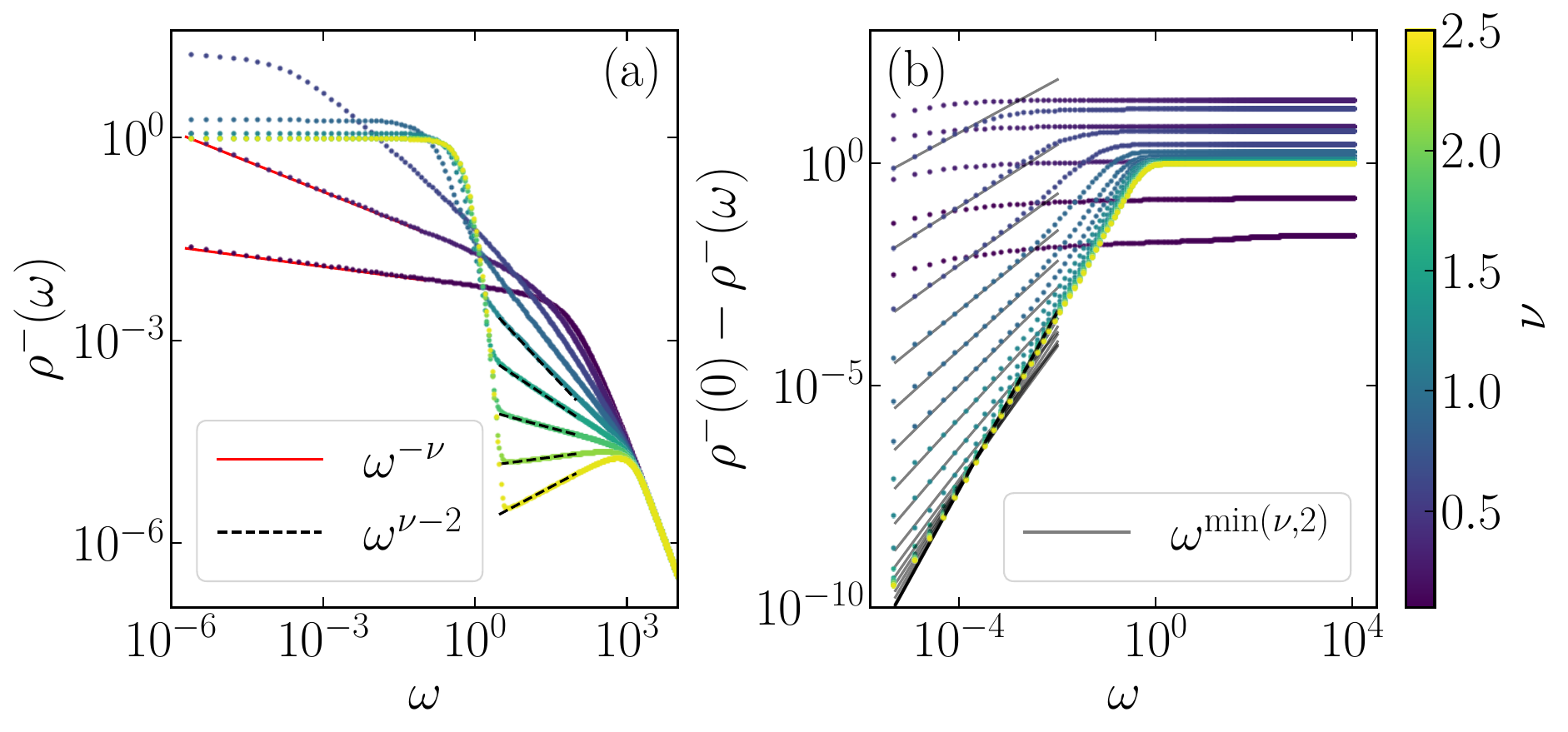}
    \caption{Self-consistent solutions of the Schwinger-Dyson equations in the strong dissipation limit for $\beta=0$, $\Lambda = 1000$, $\mu=100$, and $\nu \in [0,2.5]$ (represented by different colors). (a) $\rho^-(\omega)$ diverges as $|\omega|^{-\nu}$ for small $\nu < 1$ but there is a crossover to a plateau as $\nu$ approaches 1. For $\nu >1$, $\rho^-(\omega)$ has a power-law $|\omega|^{\nu-2}$ and a regularized $\delta$-peak. (b) $\rho^-(0)-\rho^-(\omega)$ in a log-log scale to show the nonanalytic cusp $\rho^-(\omega) \approx \rho^-(0) - c |\omega|^\nu$ as $\omega\to 0$, which for $\nu < 2$ dominates over the next-order contribution $\omega^2$.} 
    \label{fig:inf-dissipation}
\end{figure}

To gain analytical insight, we first consider the limit of strong dissipation $J\ll \Lambda \ll \mu$, where the nonlinear triple-convolution term in Eq.~(\ref{eq:sde-sigma-minus}) becomes negligible and thus $\sigma^-(\omega)=\mu D(\omega)/\pi+\epsilon$, with $\epsilon=\mathcal{O}(J)$ a constant originating from interactions. Substituting this into Eq.~\eqref{eq:sde-rho}, we obtain the low-frequency behavior of the spectral density. For $\nu < 1$,
\begin{equation}
\rho^-(\omega) \approx 
\frac{\cos^2(\pi\nu/2)}{\pi \mu}
\left|\frac{\omega}{\Lambda}\right|^{-\nu},
\qquad \text{for } \omega \lesssim \Lambda .
\end{equation}
For $\nu > 1$, the behavior depends on the frequency regime. At very low frequencies,
\begin{equation}
\rho^-(\omega) \approx
\frac{1}{\pi^2 \epsilon}
- \frac{\mu}{\pi^2 \epsilon^3}
\left|\frac{\omega}{\Lambda}\right|^\nu
- \frac{1}{\pi^4 \epsilon^3} \omega^2,
\qquad \text{for } \omega \lesssim J ,
\end{equation}
while in the intermediate regime,
\begin{equation}
\rho^-(\omega) \approx
\frac{\mu}{\pi \Lambda^2}
\left|\frac{\omega}{\Lambda}\right|^{\nu - 2},
\qquad \text{for } J \lesssim \omega \lesssim \Lambda .
\end{equation}
Each expression is valid to leading order within its respective frequency regime. Together, they reveal a sharp transition in the low-frequency behavior of the spectral function $\rho^-(\omega)$ as a function of the pseudogap exponent $\nu$. For $\nu < 1$, $\rho^-(\omega)$ diverges as $|\omega|^{-\nu}$, which produces a power-law in time with exponent $p=1+\nu$. In contrast, $\rho^-(\omega)$ is finite for $\nu >1$, with corrections of order $|\omega|^\nu$ and $\omega^2$. When $1< \nu < 2$, the nonanalytic term $|\omega|^\nu$ dominates, again producing a power-law in time with exponent $p = 1+\nu$, but for $\nu > 2$ the analytic contribution is dominant, leading to exponential relaxation at late times. Numerical solutions of the Schwinger-Dyson equations, shown in Fig.~\ref{fig:inf-dissipation}, confirm this behavior. Finally, we found that setting exactly $J= \epsilon=0$ yields a $\delta(\omega)$ contribution to $\rho^-(\omega)$ with a $\nu$-dependent prefactor; however, interactions regularize this $\delta$-function into the aforementioned behavior.

\begin{figure}
    \centering
    \includegraphics[width=0.99\linewidth]{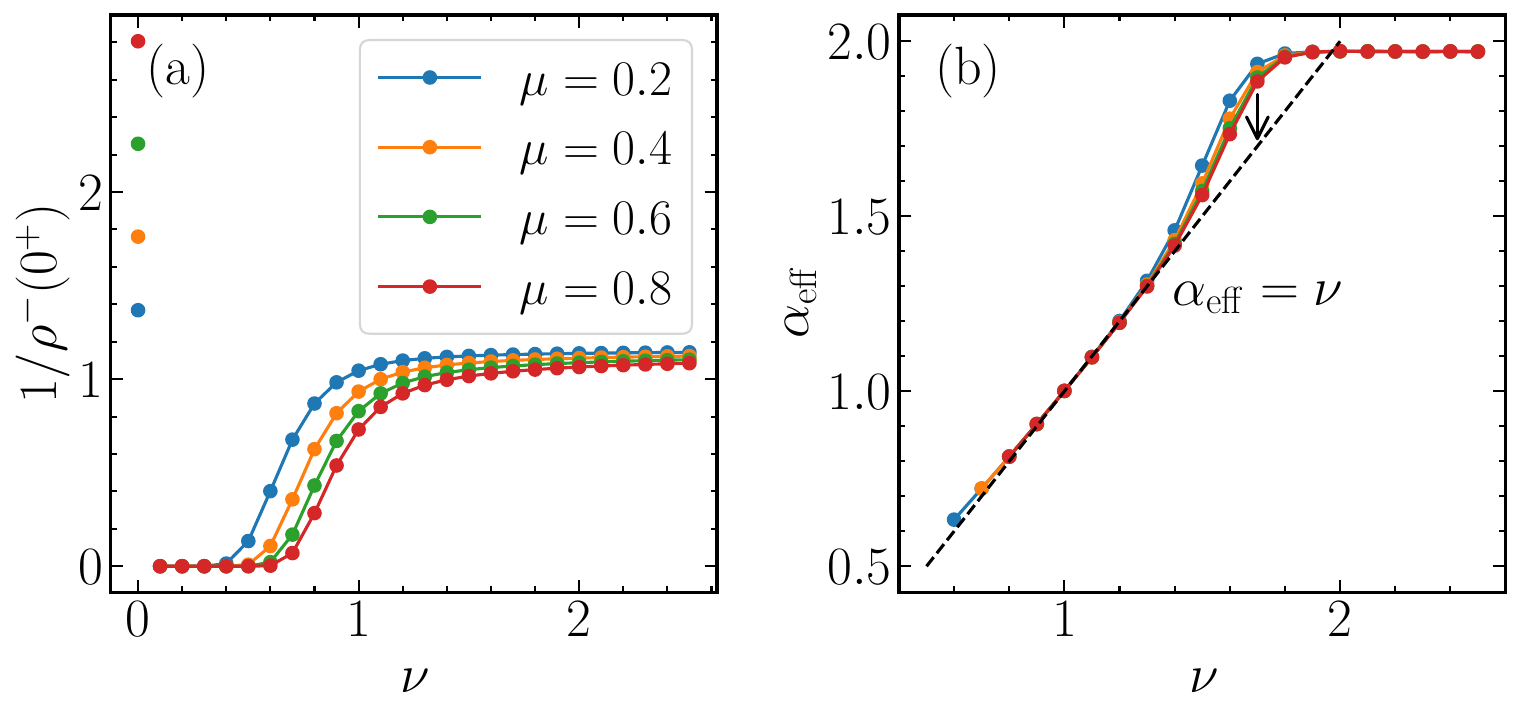}
    \caption{Characterization of the low-frequency behavior of $\rho^-(\omega)$ for $\beta=0$ and $\Lambda=10$. (a) Divergence indicator $1/\rho^-(0^+)$ as a function of $\nu$ for different values of $\mu$. The Markovian limit $\nu= 0$ is discontinuous since, unlike any small $\nu > 0$, the spectral density is finite for all $\mu$. (b) Effective exponent $\alpha_\text{eff}$ of the cusp, compared to the analytical prediction $\alpha=\nu$. Although there are some deviations as $\nu\to 2^-$, we verified that these are mitigated when the grid resolution is increased.}
    \label{fig:phase-diagram}
\end{figure}

We now show that the preceding structure persists at finite dissipation strength. To characterize the infrared behavior of $\rho^-(\omega)$, we introduce the divergence indicator $1/\rho^-(0^+)$, which vanishes when the spectral density diverges; see Fig.~\ref{fig:phase-diagram}(a). In the case of finite $\rho^-(0^+)$, we further extract the effective cusp exponent
\begin{equation}
\alpha_{\rm eff}=\lim_{\omega\to0}\frac{d\ln[\rho^-(0)-\rho^-(\omega)]}{d\ln\omega}.
\end{equation}
Assuming a low-frequency form $\rho^-(\omega)\simeq\rho^-(0)-c|\omega|^\alpha$, with $c$ a constant, and inserting it into the Schwinger-Dyson equations yields the analytical prediction $\alpha=\nu$, which is confirmed numerically in Fig.~\ref{fig:phase-diagram}(b).
We note, however, that the effective exponent $\alpha_\text{eff}$ shows deviations from the analytical predictions as $\nu \to 2^-$, originating from the finite discretization of the frequency grid. The frequency scale at which $c\,|\omega|^\nu$ dominates over $d\,\omega^2$, with $c$ and $d$ constants, is given by $\omega_\ast \sim \left(c/d\right)^{1/(2-\nu)}$, which vanishes as $\nu \to 2^-$. Although we verified that the agreement can be improved by increasing the grid resolution, any discretized approach inevitably struggles to resolve the asymptotic regime close to $\nu=2$. Remarkably, the very mechanism that handicaps the discretized approach is also responsible for a physically meaningful divergent timescale $t_\ast = 1/\omega_\ast$, which gives rise to a pre-relaxation regime with a crossover from exponential to power-law relaxation.

Combining the previous results, we arrive at a dynamical phase diagram in the $(\mu,\nu)$ plane, shown in Fig.~\ref{fig:cartoon-phase-diagram}(b), with three distinct regimes: (i) a bath-driven algebraic regime for $\nu < \nu_c(\mu)$, characterized by divergent $\rho^-(\omega)$ and power-law relaxation; (ii) an interaction-driven exponential regime for $\nu>2$, where the bath is effectively gapped and relaxation proceeds as in the isolated SYK model; and (iii) an intermediate pre-relaxation regime $\nu_c(\mu) <\nu<2$, featuring a crossover from exponential to algebraic decay, which results from a finite but nonanalytic $\rho^-(\omega)$. The phase boundary $\nu_c(\mu)$ is compatible with $\nu_c(\mu)\to1$ in the strong dissipation limit, in agreement with the analytical results above. Moreover, although our analysis focused on infinite bath temperature, we find that the dynamical phase diagram is robust against finite $\beta$, with only quantitative modifications to the crossover scales (see the SM \cite{SM}).

\begin{figure}[t]
    \centering
    \includegraphics[width = 0.99\linewidth]{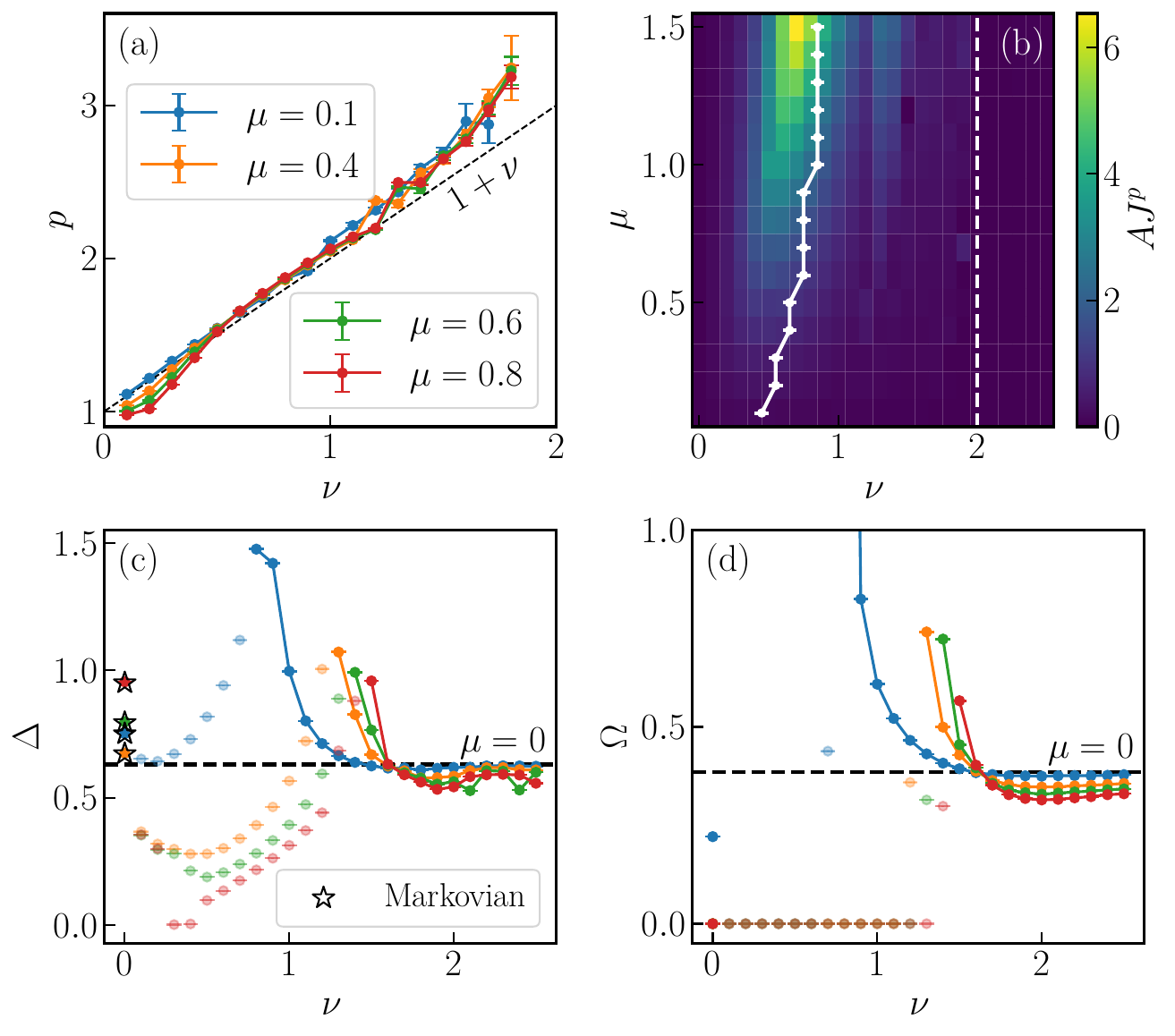}
    \caption{Characterization of the decay of $G^R(t)$ for $\beta=0$ and $\Lambda=10$, obtained by fitting the late time behavior of $iG^R(t)$ to Eq.~\eqref{eq:fit-expr}. (a) Power-law exponent $p$, which agrees with the analytical prediction $p=1+\nu$ up to deviations due to the discretization of the frequency grid. The error bars denote fit uncertainties. (b) Dimensionless parameter $AJ^p$ that quantifies power-law strength. The boundaries of the different dynamical phases, obtained from Fig.~\ref{fig:phase-diagram}, are shown in white. (c) Gap $\Delta$, which approaches its isolated-system value \cite{garcia2023keldysh} at large $\nu$. Results for $\nu=0$ from Ref.~\cite{garcia2023keldysh} are included (stars) as a benchmark and are in excellent agreement with our data. (d) Oscillation frequency $\Omega$, which likewise saturates to its isolated-system value for large $\nu$. In panels (c) and (d), shaded points correspond to fits deemed unreliable due to sensitivity to the fitting window. }
    \label{fig:time-domain}
\end{figure}

Finally, to probe the time-domain dynamics directly, we fit the late-time behavior of the numerical Green’s function to 
\begin{equation}
    i G^R_\text{fit}(t) = A /t^p + B e^{-\Delta t} \cos(\Omega t + \phi),
    \label{eq:fit-expr}
\end{equation}
which captures the coexistence of exponential and power-law relaxation. As shown in Fig.~\ref{fig:time-domain}(a), the fitted exponent closely follows the analytical prediction $p=1+\nu$ we made above. As before, small deviations appear as $\nu \to 2$, as expected from the finite discretization of the frequency grid.
Figure~\ref{fig:time-domain}(b) shows that the dimensionless amplitude $AJ^p$ is maximal in the algebraic regime, suppressed in the pre-relaxation region as $\nu$ approaches 2, and vanishes for $\nu>2$, where relaxation is purely exponential. Consistently, in the latter regime, the fitted gap $\Delta$ [Fig.~\ref{fig:time-domain}(c)] and oscillation frequency $\Omega$ [Fig.~\ref{fig:time-domain}(d)] saturate to their $\mu=0$ values~\cite{garcia2023keldysh}, indicating that the strong depletion of low-energy bath states makes the system relax as if it was isolated (i.e., due to internal chaotic dynamics).

In summary, we have studied the relaxation dynamics of an open SYK model coupled to a fermionic bath with a pseudogapped density of states, using the Keldysh formalism to obtain exact results in the large-$N$ limit. Our central finding is that the infrared structure of the environment qualitatively controls the nature of relaxation, giving rise to a rich dynamical phase diagram. Pseudogapped baths induce algebraic relaxation with a universal power-law exponent set solely by the bath, while strongly suppressed baths restore exponential decay governed by intrinsic chaotic dynamics. Between these limits, we uncovered a pre-relaxation regime characterized by a crossover from exponential to algebraic decay at late times.

These results carry implications beyond the specific model studied here. The low-energy structure of a reservoir is not merely a technical detail but a decisive factor in determining whether a strongly correlated system relaxes exponentially or algebraically. This suggests that environment engineering, in which the spectral properties of a bath are deliberately tailored, could serve as a practical tool for controlling relaxation timescales in quantum simulators and other platforms where structured reservoirs are accessible. More broadly, our findings indicate that pre-relaxation phenomena and bath-induced power laws may arise generically whenever a quantum many-body system is coupled to an environment with nontrivial low-energy structure. An interesting open direction is to solve the Schwinger-Dyson equations without assuming time-translation invariance, which would give direct access to the transient dynamics and allow one to track how the distinct relaxation regimes emerge in real time. Another important open question concerns finite-$N$ effects. Going beyond the large-$N$ limit, for instance computing $1/N$ corrections to the relaxation phase diagram we found, would give valuable insight into dissipative quantum many-body dynamics.
More broadly, exploring this interplay between non-Markovian memory effects and quantum chaos in nonequilibrium steady states remains a compelling direction for future work.

\emph{Acknowledgments.---}%
GA and PR acknowledge support by CeFEMA, Centre of Physics and Engineering of Advanced Materials, under contracts UID/04540/2025 (DOI: 10.54499/UID/04540/2025), UID/PRR/04540/2025 (DOI: 10.54499/UID/PRR/04540/2025) and UID/PRR2/04540/2025 (DOI: 10.54499/UID/PRR2/04540/2025) with the Portuguese Agência para a Investigação e Inovação AI2, and through project SCALE-QLT (DOI: 10.54499/2024.16192.PEX).
MH acknowledges support from the Deutsche Forschungsgemeinschaft under grant SFB 1143 (project-id 247310070). LS was supported by a Research Fellowship from the Royal Commission for the Exhibition of 1851.

\let\oldaddcontentsline\addcontentsline
\renewcommand{\addcontentsline}[3]{}

\let\addcontentsline\oldaddcontentsline

\clearpage


\clearpage

\setcounter{table}{0}
\renewcommand{\thetable}{S\arabic{table}}%
\setcounter{figure}{0}
\renewcommand{\thefigure}{S\arabic{figure}}%
\setcounter{equation}{0}
\renewcommand{\theequation}{S\arabic{equation}}%
\setcounter{page}{1}
\renewcommand{\thepage}{SM-\arabic{page}}%
\setcounter{secnumdepth}{3}
\setcounter{section}{0}
\renewcommand{\thesection}{S\Roman{section}}%
\setcounter{subsection}{0}
\renewcommand{\thesubsection}{\arabic{subsection}}%

\onecolumngrid
\begin{center}
    \Large{\textbf{Supplementary Information for}}\\
	\Large{\textbf{Dissipation- versus Chaos-Induced Relaxation\\in Non-Markovian Quantum Many-Body Systems}}
    \vspace{2ex}
\end{center}

{
\hypersetup{linkcolor=black}
\tableofcontents
}

\section{Derivation of effective action for the collective fields}

In this appendix, we formulate the problem using the Schwinger--Keldysh path integral and derive the effective action for the collective fields. The Keldysh generating functional is defined as
\begin{equation}
    Z = \operatorname{Tr}[\rho_{t_f}]
    = \operatorname{Tr}\!\left[ U(t_f,0)\,\rho_0\,U(0,t_f)\right],
\end{equation}
where $\rho_0$ and $\rho_{t_f}$ are the initial and final density matrices of the combined system--environment setup. Although $Z = 1$ for identical forward and backward evolutions, allowing for different evolutions on the two branches enables the computation of real-time correlation functions via functional derivatives with respect to external sources. In this sense, $Z$ plays the role of a nonequilibrium partition function. 

First, we divide the forward evolution, $U(t_f,0)$, and backward evolution, $U(0,t_f)$, into infinitesimal time steps and insert coherent-state resolutions of the identity. The result is a path-integral with action defined on the Schwinger--Keldysh contour $C$, which runs from $0 \to t_f$ (forward branch, $C_+$) and back from $t_f \to 0$ (backward branch, $C_-$). The resulting real Grassmann fields $a_i(z)$, obtained from a rotation of the complex Grassmann fields~\cite{sa2022lindbladian}, are denoted by
$a_i(z)$, $i=1,\dots,N$.
These are the coherent-state variables associated with the Majorana operators $\chi_i$ of the SYK model. Explicitly,
\begin{equation}
    a_i(z) =
    \begin{cases}
        a_i^+(t) & z=t \in C_+ \\
        a_i^-(t) & z=t \in C_-.
    \end{cases}
\end{equation}
Similarly, the bath operators $Y_i(t)$ appearing in the system--environment coupling are promoted to contour fields $Y_i(z)$. As a result, the generating functional can be written (apart from a multiplicative constant) as
\begin{equation}
    Z = \int \prod_i \mathcal{D}a_i \;
    e^{i S_\text{SYK}[a_i]}
    \left\langle
        \exp\!\left(
            i \sqrt{\mu}
            \int_C dz \sum_i a_i(z) Y_i(z)
        \right)
    \right\rangle_\mathrm{E},
\end{equation}
where the SYK action is given by
\begin{equation}
    i S_\text{SYK}[a_i] = i \int_C dz  \frac{1}{2} \sum_{i=1}^N a_i(z) i\partial_z a_i(z) - i \int_C dz \sum_{i<j<k<l}^{N} J_{ijkl} a_i(z)a_j(z)a_k(z)a_l(z),
\end{equation}
and $\langle \cdots \rangle_\mathrm{E}$ denotes a functional average weighted by the environment action, $S_\mathrm{E}[Y]$, given by
\begin{equation}
    \langle \cdots \rangle_\mathrm{E} =
    \int \prod_i\mathcal{D} Y_i (\cdots)
    e^{i S_\mathrm{E}[Y]}.
\end{equation}

Assuming $H_\mathrm{E}$ is quadratic and $\langle Y_i\rangle = 0$, we can use a cumulant expansion to write 
\begin{equation}
    \left\langle \exp\left\{i \sqrt{\mu} \int dz \sum_i a_i(z) Y_i(z)\right\}\right\rangle_\mathrm{E} = \exp \left[\frac{\mu}{2} \int dzdz' \sum_i a_i(z) a_i(z')K(z,z')\right],
\end{equation}
where $K(z,z') = - \left\langle \mathbb{T}_z Y_i(z) Y_i(z')\right\rangle_\mathrm{E}$. The last expression is exact for noninteracting baths, but a similar expression is approximately valid for other baths at leading order in $\mu$. Hence, the full action is given by
\begin{align}
    i S[a_i] = i \int_C dz  \frac{1}{2} \sum_{i=1}^N a_i(z) i\partial_z a_i(z)
    - i \int_C dz \sum_{i<j<k<l}^{N} J_{ijkl} a_i(z)a_j(z)a_k(z)a_l(z)
    +\int_C dzdz' K(z,z') \frac{\mu}{2} \sum_i a_i(z) a_i(z').
    \label{eq:full-action}
\end{align}

The dissipative term can be readily written in terms of the mean-field Green's function,
\begin{equation}
    G(z,z') = -\frac{i}{N} \sum_{i=1}^N a_i(z) a_i(z').
\end{equation}
To also write the Hamiltonian term with this collective field, we can average over the disordered couplings $J_{ijkl}$. Using the identity
\begin{equation}
    \langle e^{xJ}\rangle= e^{\frac{1}{2}\langle J^2\rangle x^2},
\end{equation}
valid for a Gaussian distribution with $\langle J\rangle=0$, we have
\begin{align}
&\left\langle \exp\left\{- i \int_C dz \sum_{i<j<k<l} J_{ijkl} a_i(z) a_j(z) a_k(z) a_l(z) \right\}\right\rangle \\
&= \exp\left\{- \frac{1}{2} \int_C dz dz' \sum_{i<j<k<l} \langle J_{ijkl}^2\rangle a_i(z) a_j(z) a_k(z) a_l(z) a_l(z') a_k(z') a_j(z') a_i(z')\right\}\\
&= \exp\left\{- \frac{1}{2} \int_C dz dz' \frac{3! J^2}{N^3} \frac{1}{4!}\left(\sum_i a_i(z) a_i(z') \right)^4 \right\}\\
&=\exp\left\{- \frac{N J^2}{8} \int_C dz dz' G(z,z')^4 \right\}.
\end{align}
We thus have that $\langle Z\rangle= \int \mathcal{D}a \exp\{i S_\text{avg}\}$, with 
\begin{equation}
    i S_\mathrm{avg}[a_i] = i\int_C dz~ \frac{1}{2} \sum_i a_i(z)i \partial_z a_i(z) - \frac{N J^2}{8} \int_C dz dz'~ G(z,z')^4 + i\frac{\mu}{2} N \int_C dzdz'~ K(z,z') G(z,z').
\end{equation}

We now make $G$ a dynamical field by inserting the identity operator
\begin{equation}
    1 \propto \int \mathcal{D}G \mathcal{D}\Sigma ~\exp \left\{ - \frac{N}{2} \int_C dzdz' \Sigma(z,z') \left(G(z,z') + \frac{i}{N} \sum_i a_i(z)a_i(z')\right)\right\}.
\end{equation}
With this, the averaged partition function becomes
\begin{equation}
    \langle Z \rangle = \int \mathcal{D}G \mathcal{D}\Sigma e^{i S_1[G,\Sigma]} \left(\int \mathcal{D}a e^{iS_2[a,\Sigma]}\right)^N,
\end{equation}
with 
\begin{equation}
    i S_1[G,\Sigma] = \frac{N}{2}\left[- \int_C dz dz'~ \Sigma(z,z') G(z,z') - \frac{J^2}{4} \int_C dz dz' G(z,z')^4 + i \mu  \int_C dz dz'~K(z,z') G(z,z')\right]
\end{equation}
and
\begin{equation}
    i S_2[a, \Sigma] = i\int_C dz~ \frac{1}{2} \sum_i a_i(z)i \partial_z a_i(z) - \frac{1}{2} \Sigma(z,z') \sum_i a_i(z)a_i(z').
\end{equation}
Crucially, $S_2$ is quadratic in the $a_i$ fields. After integration over these fields, we get the final effective action 
\begin{align}
    i S_\text{eff}[G,\Sigma] = \frac{N}{2} &\left\{ \operatorname{Tr} \log (i\partial - \Sigma ) - \int_C dzdz' \Sigma(z,z') G(z,z') - \frac{J^2}{4} \int_C dz dz' G(z,z')^4 \right. \notag \\
     &\left. + i \frac{\mu}{2} \int_C dzdz' K(z,z') G(z,z')\right\}.
     \label{eq:S_eff}
\end{align}

\section{Non-Markovian Schwinger-Dyson equations}

In this appendix, we derive and discuss the properties of the non-Markovian Schwinger-Dyson equations. While in the Main Text [Eqs.~\eqref{eq:sde-sigma-minus}--\eqref{eq:sde-rho}] we specialized to $\beta=0$, here we consider the general case.
The action in Eq.~\eqref{eq:S_eff} is proportional to the number of fermions $N$. Therefore, in the large-$N$ limit, the path-integral is dominated by the saddle point of $S_\text{eff}[G,\Sigma]$. Requiring stationarity with respect to variations of $G$ and $\Sigma$ yields the Schwinger--Dyson equations
\begin{align}
    &(i\partial - \Sigma) \cdot G = \mathbbm{1}_C, \label{eq:sde1}\\
    &\Sigma(z,z') = - J^2 G(z,z')^3 + i\frac{\mu}{2} \left( K(z,z') - K(z',z) \right),\label{eq:sde2}
\end{align}
where the Eq.~\eqref{eq:sde1} is to be understood as a matrix equation in the space of contour times $(z,z') \in C^2$, and $\mathbbm{1}_C$ denotes the identity on the contour.

To make contact with real-time correlation functions, we project the contour variables
$z,z' \in C$ onto the forward $(+)$ and backward $(-)$ branches of the Keldysh contour. Concretely,
we set $z = t_1^\pm$ and $z' = t_2^\mp$ in Eq.~\eqref{eq:sde2}, where the superscripts $\pm$ indicate
whether the time argument lies on the forward or backward branch. This allows us to identify the
greater and lesser components of the self-energy,
\begin{align}
    \Sigma^>(t_1,t_2) &\equiv  \Sigma(t_1^+,t_2^-) = -J^2 G^>(t_1,t_2)^3 + i\frac{\mu}{2} \left[K^>(t_1,t_2) - K^<(t_2,t_1)\right],\\
    \Sigma^<(t_1,t_2) &\equiv  \Sigma(t_1^-,t_2^+)= -J^2 G^<(t_1,t_2)^3 + i\frac{\mu}{2} \left[K^<(t_1,t_2) - K^>(t_2,t_1)\right].
\end{align}
Here, the greater and lesser Green’s functions and memory kernels are defined by $G^{>(<)}(t_1,t_2) \equiv G(t_1^{+(-)},t_2^{- (+)})$ and $K^{>(<)}(t_1,t_2) \equiv K(t_1^{+(-)},t_2^{- (+)})$, respectively. For Majorana fermions, the greater and lesser Green’s functions are not independent but satisfy $G^<(t_1,t_2) = - G^>(t_2,t_1)$, which considerably simplifies the structure of the equations.

We now introduce ``center-of-mass'' and relative time coordinates, $\tau = (t_1+t_2)/2$ and $t=t_1 - t_2$. In the long-time limit $\tau \to \infty$ (which requires letting $t_f \to \infty$), the system relaxes to the steady state $\rho_\infty$, and all correlation functions become time-translational
invariant, depending only on the relative time $t$. This allows us to Fourier transform the Schwinger--Dyson equations, yielding
\begin{align}
    \Sigma^>(\omega) = &-\frac{J^2}{(2\pi)^2} \left(G^> \ast G^>
     \ast G^>\right)(\omega) - i\mu \frac{e^{\beta \omega}}{1 + e^{\beta \omega}} D(\omega), \\
    \Sigma^<(\omega) = &-\frac{J^2}{(2\pi)^2} \left(G^< \ast G^<
     \ast G^<\right)(\omega) + i\mu \frac{1}{1 + e^{\beta \omega}} D(\omega),
\end{align}
where $\ast$ denotes convolution in frequency space. 
In deriving these expressions, we have used the fluctuation--dissipation relations satisfied by the
bath correlation functions, which imply that the memory kernel components
are given by
\begin{align}
    K^>(\omega) &= -\bigl(1 - n_F(\omega)\bigr) D(\omega), \\
    K^<(\omega) &= \phantom{-}n_F(\omega)\, D(\omega),
\end{align}
where $n_F(\omega) = \left(1 + e^{\beta\omega}\right)^{-1}$ is the Fermi-Dirac distribution. For an infinite-temperature bath with a constant density of states, both $n_F(\omega)$ and $D(\omega)$ become independent of $\omega$ and thus the memory Kernel becomes a $\delta$-function in the time domain. This corresponds to the Markovian limit, since the action in Eq.~\eqref{eq:full-action} becomes local in time. In fact, the resulting action could have been obtained from a Lindblad evolution $\dot{\rho} = \mathcal{L}[\rho]$ with jump operators given by $L_i = \sqrt{\mu} \chi_i$.

Finally, we perform a Keldysh rotation by defining the real quantities
\begin{align}
    &\rho^\pm (\omega) = - \frac{1}{2\pi i} \left(G^>(\omega) \mp G^>(-\omega)\right),\\
    &\rho^H (\omega) = - \frac{1}{\pi} \mathcal{P} \int d\omega' \frac{\rho^-(\omega')}{\omega - \omega'},
\end{align}
where $\mathcal{P}$ denotes the Cauchy principal value. Analogous definitions are made for $\sigma^\pm(\omega)$ and $\sigma^H(\omega)$ in terms of $\Sigma^>(\omega)$. 

With these new quantities, Eq.~\eqref{eq:sde2} is equivalent to 
\begin{align}
    \sigma^+(\omega) &= \frac{J^2}{4} \left(\rho^+ \ast \rho^+ \ast \rho^+  + 3 \rho^- \ast \rho^- \ast \rho^+ \right)(\omega) + \frac{\mu}{\pi} \tanh\left(\frac{\beta \omega}{2}\right) D(\omega), \label{sde:sigma-plus-finiteT} \\
    \sigma^-(\omega) &= \frac{J^2}{4} \left(\rho^- \ast \rho^- \ast \rho^-  + 3 \rho^+ \ast \rho^+ \ast \rho^- \right)(\omega) + \frac{\mu}{\pi} D(\omega), \label{sde:sigma-minus-finiteT}
\end{align}
while Eq.~\eqref{eq:sde1} is given by
\begin{equation}
    \rho^\pm (\omega) = \frac{\sigma^\pm (\omega)}{(\omega  + \pi \sigma^H(\omega))^2 + (\pi \sigma^-(\omega))^2}.
    \label{eq:sde-rhos-pm}
\end{equation}
These are the Schwinger-Dyson equations that determine the steady-state Green's function. For $\beta = 0$, inspection of the equations implies that $\rho^+(\omega)=\sigma^+(\omega)=0$, which is also expected from considering the fluctuation-dissipation relation, Eq.~\eqref{eq:fdr-rhos}. The resulting equations are Eqs.~\eqref{eq:sde-sigma-minus}--\eqref{eq:sde-rho} in the Main Text.

\section{Numerical methods}

In this appendix, we discuss how to numerically solve the Schwinger-Dyson equations [Eqs.~\eqref{sde:sigma-plus-finiteT}, \eqref{sde:sigma-minus-finiteT}, and ~\eqref{eq:sde-rhos-pm}] self-consistently. We start by discretizing the frequencies into a grid that accumulates exponentially at $\omega = 0$ and has a regular spacing $\Delta \omega$ for $\omega\gtrsim  \Lambda$. For this purpose, we use the parametrization $\omega = \pm x e^{-x_0/x}$ with $x$ taken uniformly between $x_\text{min}$ and $x_\text{max}$ with regular spacing $\Delta x$. This parametrization ensures an exponential accumulation at $\omega = 0$ that smoothly transitions at $x_0$ to a regularly spaced grid with spacing $\Delta \omega = \Delta x$. We take $x_0 = 5 \Lambda$, $x_\text{min} = 0.27\Lambda$, $x_\text{max} = 15 \Lambda$ and $\Delta x \in \{0.02,0.01\} \Lambda$ (recall that, except for Fig.~\ref{fig:inf-dissipation}, $\Lambda=10J$). 

Then, we iteratively compute $\sigma_{i+1}^\pm$ from $\rho^\pm_i$, using Eq.~\eqref{eq:sde1}, and do a partial update
\begin{equation}
    \rho^\pm_{i+1}(\omega) = (1-\eta) \rho_i^\pm(\omega) + \eta \frac{\sigma_{i+1}^\pm(\omega)}{(\omega + \pi \sigma_{i+1}^H(\omega))^2 + (\pi \sigma_{i+1}^-(\omega))^2},
\end{equation}
with $\eta = 0.1$, until we reach the fixed point (i.e., self-consistency), meaning that
\begin{equation}
    \sum_{n} \frac{|\rho_{i+1}^+(\omega_n) - \rho_{i}^+(\omega_n)| +  |\rho_{i+1}^-(\omega_n) - \rho_{i}^-(\omega_n)|}{2\eta} < \epsilon_\text{tol} = 10^{-3}.
\end{equation}
The partial update with small $\eta$ is used to ensure monotone convergence to the fixed point. 

As a benchmark for consistency and to speed up convergence, we use two frequency grid spacings $\Delta \omega \in \{0.2,0.1\} J$. We start by obtaining solutions for the coarser grid, while for the finer grid, we initialize $\rho^\pm$ to an interpolation of the converged solution with the coarser grid. To further reduce the number of iterations, when the parameters $(\mu, \nu, \beta)$ are swiped over a range, we use the previous converged solution as an initialization of $\rho^\pm(\omega)$.

To move back to the time domain and compute $G^R(t)$, we interpolate $\rho^-(\omega)$ into a regular frequency grid with $\Delta \omega'=10^{-3} J $ and apply a fast Fourier transform. After obtaining $G^R(t)$, we perform a power-law fit to the late-time regime, defined as the final 30\% of the available data points. The resulting fit parameters are used as an initial guess for the full fit combining exponential and power-law decay, Eq.~\eqref{eq:fit-expr}, performed for $Jt>4$.

\section{Finite-temperature relaxation results}

In this appendix, we provide evidence that the qualitative results found in the Main Text still hold at finite bath temperature ($\beta\neq 0$). Numerically, we find that the converged $\rho^\pm$ approximately satisfy the fluctuation-dissipation relation, Eq.~(\ref{eq:fdr-rhos}), implying that the system equilibrates with the bath at inverse temperature $\beta$. 

\begin{figure}[b]
	\centering
    \includegraphics[width=0.5\linewidth]{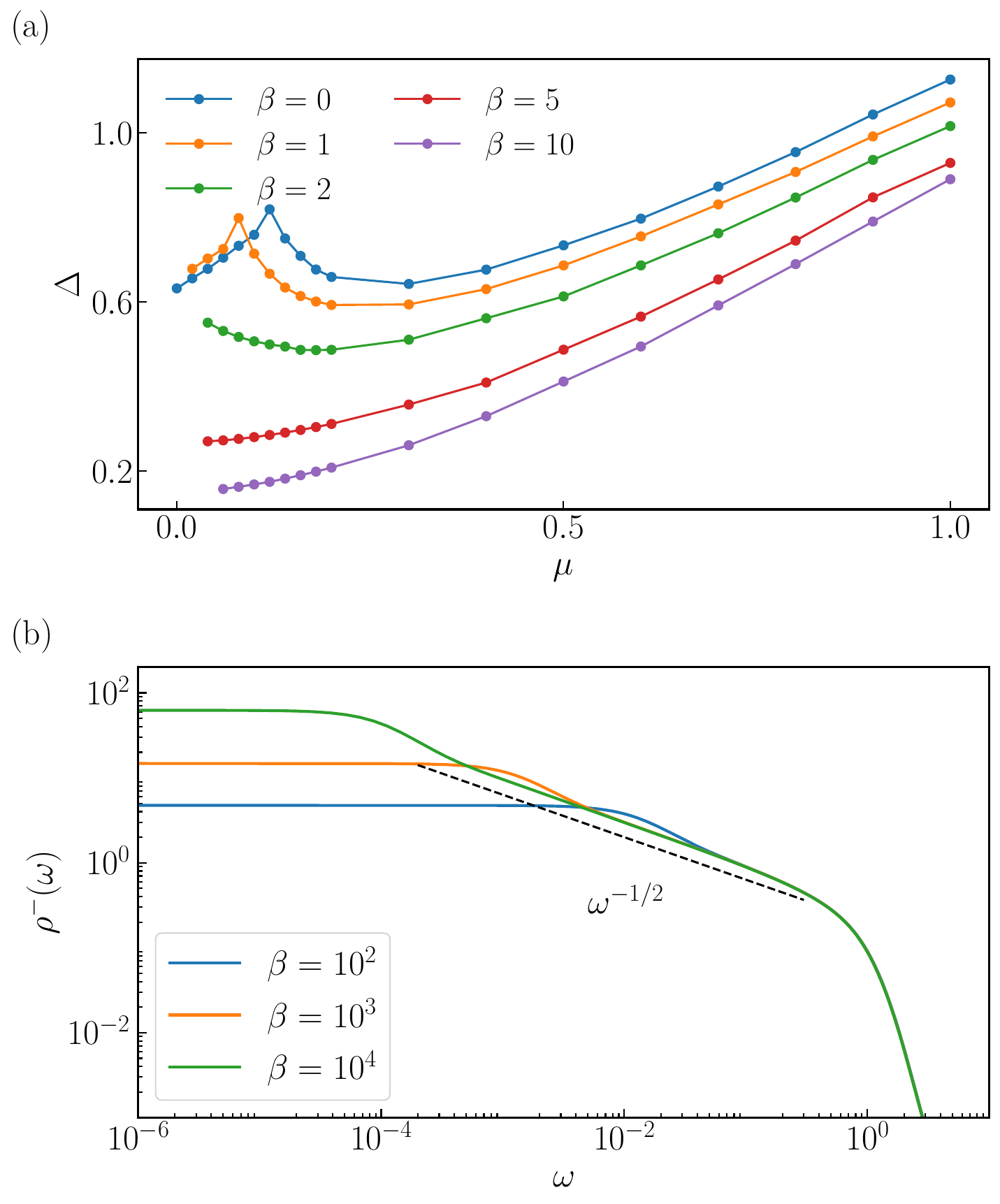}
	\caption{Finite-temperature relaxation for $\nu=0$ (i.e., an environment with a constant density of states). (a) Gap $\Delta$ as a function of $\mu$ and $\beta$. (b) $\rho^-(\omega)$ behavior in the low-temperature limit. As $\mu \to 0$, $\rho^-(\omega) \sim \omega^{-1/2}$ up to $\omega \approx 1/\beta$. For this panel only, we impose the fluctuation-dissipation relation, Eq.~\eqref{eq:fdr-rhos}, at every iteration to make the convergence feasible. 
    }
	\label{fig:finite-temp-nu0}
\end{figure}

We first consider $\nu=0$, meaning that the environment has a constant density of states. At infinite bath temperature, the dissipation is Markovian and the system relaxes exponentially. Although a finite temperature bath gives rise to non-Markovian dissipation, we find that $G^R(t)$ still decays exponentially. However, in the zero-temperature limit ($\beta \to \infty$), the isolated system ($\mu = 0$) has a nonanalytic spectral density $\rho^-(\omega) \sim \omega^{-1/2}$ as $\omega\to 0$, due to the emergent conformal symmetry. Consequently, the retarded Green's function $iG^R(t)\sim t^{-3/2}$ as $t \to +\infty$ \cite{maldacena2016remarks}. Thus, as we approach this limit, the gap $\Delta$ vanishes and $\rho^-(\omega)\sim\omega^{-1/2}$ up to the scale $\omega\approx 1/\beta$, where the system ``sees'' the finite temperature of the bath; see Fig.~\ref{fig:finite-temp-nu0}.

\begin{figure}
	\centering
	\includegraphics[width=0.7\linewidth]{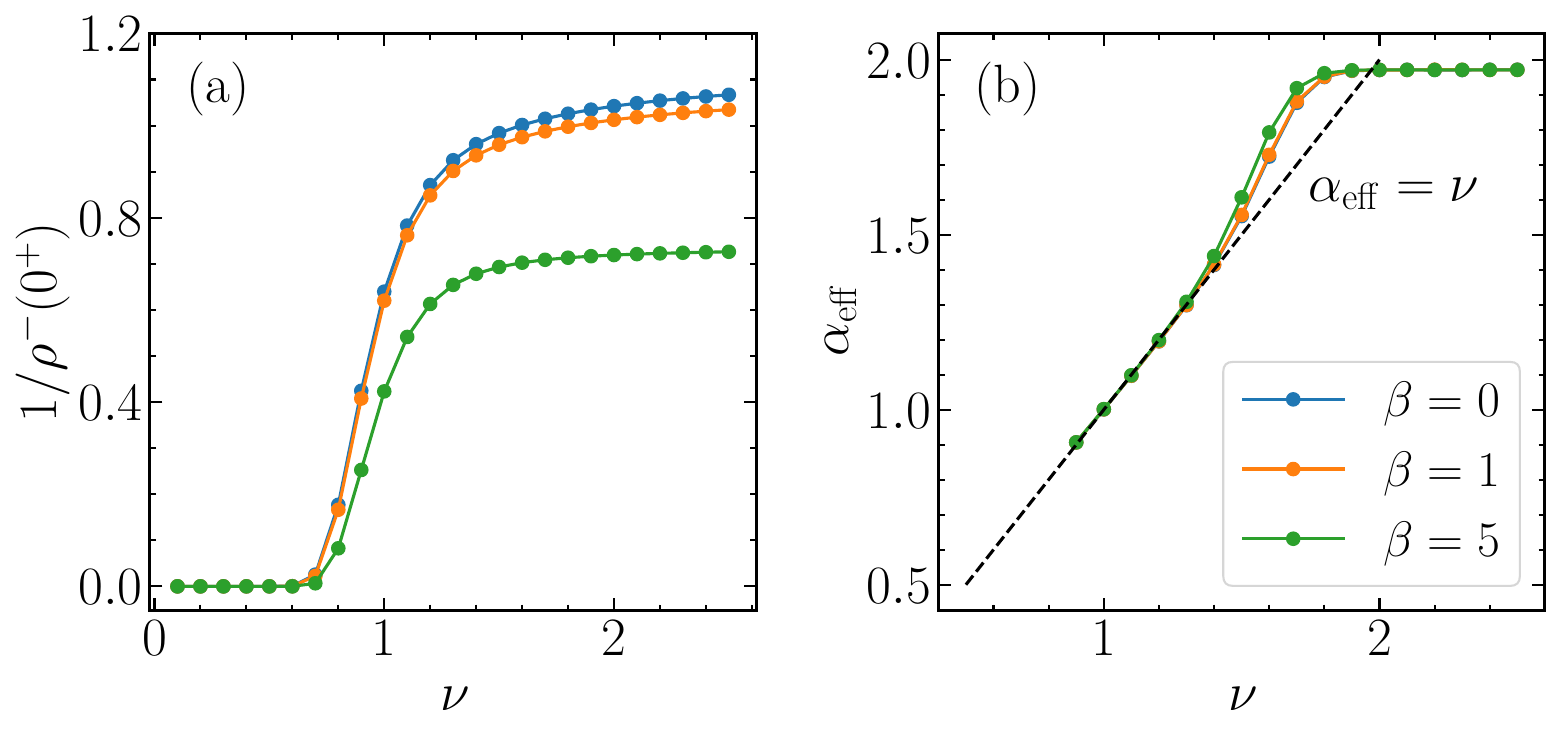}
	\caption{Characterization of the low-frequency behavior of $\rho^-(\omega)$ for three inverse bath temperatures $\beta \in \{0, 1, 5\}$. We show the divergence indicator $1/\rho^-(0^+)$ (a) and the effective cusp exponent $\alpha_\text{eff}$ (b) as a function of $\nu$ for $\mu = 1$ and $\Lambda = 10$. This corresponds to a horizontal stripe of the phase diagram of Fig.~\ref{fig:cartoon-phase-diagram}(b) but at finite temperature, showing that the three-regime structure is robust to finite bath temperature. Here, after verifying that the solutions still respect fluctuation-dissipation relations for a few points, we impose them systematically at every iteration step. } 
	\label{fig:finite-temp-pseudogap}
\end{figure}

For the pseudogapped environment ($\nu>0$), we analyze a stripe of the phase diagram at $\mu = 1$ for two finite bath temperatures. The results, shown in Fig.~\ref{fig:finite-temp-pseudogap}, demonstrate that the same three-regime structure is found: a divergent spectral density for small $\nu$; a nonanalytic cusp for intermediate $\nu$ that competes with the analytic contribution $\omega^2$, producing the pre-relaxation behavior; and a region where the nonanalytic contribution is subdominant, restoring exponential relaxation.

\end{document}